\documentclass[letterpaper,twocolumn,aps,prb,showpacs]{revtex4}
\usepackage{graphicx}
\usepackage{amsmath}
\usepackage{amsfonts}
\usepackage{amssymb}
\usepackage{epstopdf}

\begin{document}
\author{Yaroslav Tserkovnyak}
\affiliation{Lyman Laboratory of Physics, Harvard University, Cambridge, Massachusetts 02138, USA}
\affiliation{Department of Physics and Astronomy, University of California, Los Angeles, California 90095, USA}
\author{Bertrand I. Halperin}
\affiliation{Lyman Laboratory of Physics, Harvard University, Cambridge, Massachusetts 02138, USA}

\title{Magnetoconductance oscillations in quasiballistic multimode nanowires}

\begin{abstract}
We calculate the conductance of quasi-one-dimensional nanowires with electronic states confined to a surface charge layer, in the presence of a uniform magnetic field. Two-terminal magnetoconductance (MC) between two leads deposited on the nanowire via tunnel barriers is dominated by density-of-states (DOS) singularities, when the leads are well apart. There is also a mesoscopic correction due to a higher-order coherent tunneling between the leads for small lead separation. The corresponding MC structure depends on the interference between electron propagation via different channels connecting the leads, which in the simplest case, for the magnetic field along the wire axis, can be crudely characterized by relative winding numbers of paths enclosing the magnetic flux. In general, the MC oscillations are aperiodic, due to the Zeeman splitting, field misalignment with the wire axis, and a finite extent of electron distribution across the wire cross section, and are affected by spin-orbit coupling. The quantum-interference MC traces contain a wealth of information about the electronic structure of multichannel wires, which would be complimentary to the DOS measurements. We propose a four-terminal configuration to enhance the relative contribution of the higher-order tunneling processes and apply our results to realistic InAs nanowires carrying several quantum channels in the surface charge-accumulation layer.
\end{abstract}
\date{\today}
\maketitle

\section{Introduction}
Magnetoconductance (MC) of very small conductors, e.g., quantum dots, wires, and rings, is a topic of intense research since the early days of mesoscopic physics. In the simple case of phase-coherent conductors topologically identical to the ring, one can generally distinguish between two classes of MC oscillations, differing in frequency by the factor of $2$. First of all, the conductance oscillates with a period of $h/e$ in magnetic flux (Aharonov-Bohm effect), since the magnetic field through the loop can be removed by a gauge transformation in integer multiples of the flux quantum $h/e$. These oscillations are sensitive to the disorder distribution, however, so that averaging over various disorder configurations in nominally identical conductors, or self-averaging in larger samples, reduces the strength of the oscillations. The quantum weak-localization (or antilocalization) correction with a period of $h/2e$ survives the averaging in diffusive samples.\cite{altshulerJETPL81} Depending on the concentration of the disorder in a given sample, either $h/e$- or $h/2e$-type oscillations can be more pronounced. A given realization of the disordered phase-coherent mesoscopic conductor, such as an open quantum dot or quasi-one-dimensional (1D) wire, also exhibits universal conductance fluctuations (UCF) with amplitude $\sim e^2/h$, which vanish on sample averaging.\cite{beenakkerRMP97} The weak-localization correction related to $h/2e$ MC oscillations in rings is usually manifested as a smooth background with the superimposed aperiodic UCF oscillations.

All of the aforementioned phenomena can play a role in a quasi-1D wire with a finite cross section penetrated by a magnetic field. While a significant volume of the existing literature is devoted to the MC studies of disordered wires (see, e.g., Refs.~\onlinecite{stonePRL85} and \onlinecite{hansenPRB05}), the interference effects in magnetotransport of the quasiballistic nanowires did not attract the same attention (see, however, Refs.~\onlinecite{dugaevJETP84} for a discussion of weak-localization effects in narrow two-dimensional wires, which are ballistic across the wire width). Recent advances in nanofabrication techniques, however, make it possible to experimentally probe the MC of nearly ballistic semiconducting nanowires,\cite{lieberPRIV02} raising questions about the conductance oscillations related to the finite separation of the leads. See also Ref.~\onlinecite{liangNAT01}, where ballistic carbon nanotubes were used to build a Fabry-P{\'e}rot electron resonator, and Refs.~\onlinecite{tserkovPRL02hw}, reporting an observation and theory of the interference effects in tunneling between parallel ballistic quantum wires through a finite-length barrier.

In this paper, we are studying theoretically the magnetotransport in single quasiballistic nanowires. Suppose the wire carries more than one quantum channel, so that the electron propagation between the injecting and detecting leads can interfere giving rise to conductance oscillations in the magnetic field, which depend on the lead separation length. If the electron density is confined to a narrow surface layer and the Zeeman energy is disregarded, the MC in the parallel magnetic field can be decomposed into harmonics with frequencies in multiples of $h/e$ approximately corresponding to relative winding numbers of various electron paths around the wire axis between the two leads. Generally, different transverse modes will effectively have different cross sections for accumulating the longitudinal magnetic flux resulting in aperiodic interference patterns. The aperiodicity is further enhanced by Zeeman energy and, furthermore, the spin degrees of freedom affect transport nontrivially due to the spin-orbit coupling. In addition, strong magnetic fields can have a significant effect on the one-dimensional band structure of a long wire, leading to van Hove magnetofingerprints in the conductance. Varying the direction of the magnetic field with respect to the wire axis gives an additional degree of freedom for probing electronic structure. For example, a large perpendicular field eventually leads to a formation of Landau levels having a profound effect on the one-dimensional conductance.

Among the principal practical questions are how the MC traces can be used to extract information about electronic states in the transverse direction (such as th e total number of modes and charge distribution, e.g., surface-confined or nearly uniform etc.) and about various scattering mechanisms (involving in general elastic, phase-relaxation, and spin-orbit processes). While giving comprehensive answers to these questions lies beyond the scope of this paper, our analysis can be used to consider them in some simple scenarios as well as to lay the grounds for more systematic studies.

This work was motivated, in large part, by magnetoconductance studies\cite{lieberPRIV02} of cylindrical conducting nanowires,\cite{lieberMRSB03} which provide a fascinating medium for mesoscopic physics\cite{hansenPRB05,dohSCI05,bjorkNL04} as well as show a great promise for applications as building blocks of nanowire networks.\cite{huangSCI01,lieberMRSB03} After a general theoretical discussion in Sec.~\ref{theory}, in Sec.~\ref{results} we treat a specific case of transport in a surface charge-accumulation layer of cylindrical InAs nanowires as those studied in Ref.~\onlinecite{lieberPRIV02}, using realistic parameters extracted from photoelectron spectroscopy\cite{olssonPRL96} and magnetotransport measurements\cite{matsuyamaPRB00} on two-dimensional electron gases at InAs surfaces. Our approach may, however, also be useful for other types of conducting nanowires, such as state-of-the-art semiconductor axial heterostructures with radially modulated composition and doping.\cite{lauhonNAT02} Concluding remarks are given in Sec.~\ref{summary}.

\section{Theory}
\label{theory}

\begin{figure}
\includegraphics[width=0.8\linewidth,clip=]{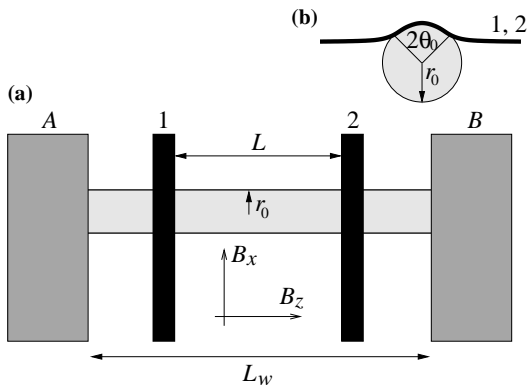}
\caption{\label{sc}Schematic of the model [view parallel to the wire (a) and of its cross section (b)]: A long wire of length $L_w$ is contacted via tunnel barrier by two leads 1 and 2 separated by distance $L$. In addition, the wire is {\O}hmically attached at the ends to two ``sinks" $A$ and $B$. Leads 1 and 2 can be voltage biased and $A$ and $B$ can be either grounded or left electrically floating. Sinks $A$ and $B$ are assumed to form good contacts with the wire and are used to either decohere (when floating) or sink (when grounded) incoming electrons, as explained in the text.}
\end{figure}

Consider a long 1D wire connected through tunnel barriers to two narrow metallic source and drain leads separated by distance $L$, as shown in Fig.~\ref{sc}. In addition, the wire may be attached to two sinks (reservoirs) $A$ and $B$, at the ends, whose purpose will become clear later. Disregarding electron-electron interactions, the wire supports $N$ transverse modes, the value of its two-terminal conductance being therefore bounded by $Ne^2/h$, at a fixed magnetic field (supposing the sinks at the ends are electrically floating, for the moment). At a finite voltage bias, electrons are injected from the left lead 1 into the wire where they decompose into different transverse modes, are carried along the wire, and are detected by the right lead 2 or disappear in the reservoirs $A$ and $B$ at the ends of the wire. If the reservoirs are floating, an electron that enters the reservoir will eventually reenter the wire, but will have lost all coherence with the original injection process. The magnetic field applied either parallel or perpendicular to the wire affects interference between different electron trajectories connecting the contacts, which accumulate the phase by enclosing a magnetic flux, resulting in conductance oscillations. In addition, the field can modify the number of quantum channels $N$ at the Fermi level resulting in van Hove, i.e., density-of-states (DOS), singularities.

We develop a simple formalism to capture interference between electrons injected into different transverse modes. Wave-function decomposition at the injecting lead is determined by the angular distribution around the wire circumference of the tunneling amplitude formed at the contact. (In a specific model discussed below, we take the tunneling strength to be uniform within an angle $2\theta_0$ determined by the lead deposition [see Fig.~\ref{sc}(b)].) For simplicity, we assume that the contacts are narrow along the wire in comparison to their separation $L$ and to the electron wavelength, and we focus on MC oscillations due to a finite length $L$. Similarly, the electrons are detected by the drain lead whose coupling to different transverse modes is determined by the deposition. We will focus on the MC oscillations due to the interference of phases accumulated during quasiballistic propagation between the leads. In particular, we will disregard any systematic contributions due to weak-localization (or antilocalization) physics, restricting our attention to oscillation features on scales determined by relative phases accumulated for multimode ballistic propagation along the length $L$. This physical picture is made mathematically concrete in the following.

\subsection{General considerations}
\label{gc}

A schematic of our model is shown in Fig.~\ref{sc}. A long wire of length $L_w$ and radius $r_0$ is contacted via tunnel barriers by two metallic leads 1 and 2 separated by distance $L$ and deposited on top of the wire [see Fig.~\ref{sc}(b)]. The total capacitance of the wire and/or the attached reservoirs is assumed large enough so that Coulomb-blockade effects can be neglected. However, $L$ may be smaller than the coherence length, and we are interested in interference effects for electron propagation between contacts 1 and 2.

Electron flow in the wire is carried by a surface charge layer. The disorder (due to impurities or wire-shape imperfections) is assumed to be weak enough, so that the mean free path $\lambda>2\pi r_0$ and it makes sense to define quasi-1D bands along the wire. If, furthermore, the lead separation $L$ is not much longer than $\lambda$, some electrons can propagate ballistically between the leads, giving rise to a magnetoresistance structure discussed in the following.

We can distinguish between several regimes based on the relation between the contact separation $L$, mean-free path $\lambda$, and the dephasing length $l_\phi$. Let us assume that $l_\phi\gg\lambda$, which can be achieved at low temperatures, and suppose the sinks $A$ and $B$ are floating. If $L\gg l_\phi\gg\lambda$, the two-terminal conductance $G_{12}$ is determined by summing three resistances in series (unaffected by $A$ and $B$),
\begin{equation}
G^{-1}_{12}=G^{-1}_1+G^{-1}_2+G^{-1}_L\,,
\label{R12}
\end{equation}
where $G_l$ is the tunneling conductance of the $l$th lead and $G_L\propto 1/L$ is the diffusive conductance of the wire section between the leads. $G_1$ can be experimentally obtained by measuring the two-terminal conductance between 1 and $A$ (while keeping the other contacts floating) and $G_2$|between 2 and $B$, for example, if the conductances of the sinks are large, $G_1,G_2\ll G_A,G_B$. The diffusive contributions from the wire can be accounted for by making additional two-terminal measurements between 1 and $B$ and 2 and $A$, if the intercontact separations are known. (We are, however, interested in the clean limit, where the diffusive resistance of the wire is small.) If $l_\phi\gg L\gg\lambda$, Eq.~(\ref{R12}) is still valid but the diffusive conductance $G_L$ now acquires mesoscopic fluctuations that can be modulated by a nearby gate or the magnetic field. We will focus on the regime $l_\phi\gg\lambda\gtrsim L$, where the electron propagation between the contacts is phase coherent and nearly ballistic.

To the lowest order in tunneling, the lead conductances $G_l$, $l=1,2$, are determined by the density-of-states of the quasi-1D bands in the wire and the respective tunneling amplitudes $T_{li}$. For the $i$th 1D mode in the wire, lead-$l$ tunneling conductance is $G_{l i}=|T_{li}|^2/v_{gi}$ (with all the trivial prefactors including the lead DOS lumped into $T_{li}$), where
\begin{equation}
v_{gi}=\partial_k\epsilon_i(k)
\label{vk}
\end{equation}
is the Fermi-point group velocity corresponding to the $i$th mode dispersion $\epsilon_i(k)$ (recall that the DOS is proportional to the inverse group velocity $1/v_{gi}$ in 1D). The tunneling conductance thus exhibits van Hove singularities when the Fermi level crosses a band edge. If electrons are narrowly confined near the surface and are restricted to the lowest band in the radial direction, tunneling amplitudes will be approximately the same for different 1D bands and thus
\begin{equation}
G_l=|T_l|^2\sum_iv_{gi}^{-1}\,.
\label{gk}
\end{equation}
In this simple model, the two-terminal conductance $G_{12}$ given by Eq.~(\ref{R12}) thus exhibits van Hove singularities whose magnetic fingerprints could be a useful tool for investigating the electronic structure of the nanowires. $G_L$ can also depend on the magnetic field, due to the weak-localization correction to semiclassical diffusion, but this will be disregarded assuming the wire is sufficiently clean.

When the lead separation is decreased and becomes comparable to the mean-free path $\lambda$, a mesoscopic correction to the conductance $\delta G_{12}$, which is higher order in tunneling, may also contribute to the magnetoresistance traces for not too opaque tunnel barriers. $\delta G_{12}$ is determined by a coherent tunneling between the leads, so it is fourth order in tunneling amplitude at the contacts and quadratic in the electron propagator in the wire,
\begin{equation}
\delta G_{12}\propto\sum_{\mu\nu}\left|\sum_{ij}M_{2,i\mu}^\ast M_{1,j\nu}\mathcal{G}_{ij}(L,E_F)\right|^2\,,
\label{green}
\end{equation}
in terms of the retarded Green's function
\begin{equation}
\mathcal{G}_{ij}(L,E_F)=\int_0^\infty dte^{iE_Ft}\langle\{\Psi_i(L,t),\Psi^\dagger_j(0,0)\}\rangle
\label{Gij}
\end{equation}
for quasi-1D electron propagation between leads $1$ (source) and $2$ (drain) at the Fermi energy $E_F$. (We set $\hbar=1$.) $M_{l,i\mu}$ is the tunneling matrix element into the $i$th channel in the wire from the lead-$l$ state labeled by $\mu$ at the Fermi energy. If the interband scattering in the wire is disregarded, Eq.~(\ref{green}) becomes
\begin{equation}
\delta G_{12}\propto\sum_{\mu\nu}\left|\sum_i M_{2,i\mu}^\ast M_{1,i\nu}\mathcal{G}_i(L,E_F)\right|^2\,.
\label{g}
\end{equation}
At low bias and vanishing temperature, all quantities entering Eqs.~(\ref{green}) and (\ref{g}) are evaluated at the Fermi energy since we are considering the elastic contribution to the tunneling. Although we will assume this in the following for simplicity, the generalization to the finite temperature and bias is straightforward for noninteracting electrons. (The temperature must, however, be at least larger than the level spacing in the wire determined by $L_w$, since we do not intend to analyze individual levels.) An additional inelastic contribution to the higher-order tunneling (when one electron is injected into the wire but a different one is extracted, leaving behind an electron-hole pair excitation) may become important at finite temperature and/or bias,\cite{averinPRL90} but is disregarded here since it does not contribute to the discussed interference structure. For simplicity, we have assumed narrow leads, so that the propagator (\ref{Gij}) is evaluated for a fixed distance $L$. The presence of sinks $A$ and $B$ at the wire ends decoheres electrons reflected from the ends and we can neglect their contribution to the propagator (\ref{Gij}).

If the incoherent contribution to tunneling determined by Eq.~(\ref{R12}) is much larger than the correction (\ref{green}) given by higher-order tunneling processes, the magnetoconductance will be dominated by the van Hove singularities, according to Eq.~(\ref{gk}) for the DOS-dependent contact conductances. It is however possible to suppress the incoherent contribution to the signal by grounding sinks $A$ and $B$ at the ends of the wire and measuring the current in a grounded contact 1 or 2, while the other contact is voltage biased. The incoherent part of the conductance corresponding to the measured current will then be given by 
\begin{equation}
G_{12}=\frac{G_1G_2}{G_1+G_2+G_A+G_B}\approx\frac{G_1G_2}{G_A+G_B}\,,
\label{gsd}
\end{equation}
neglecting for simplicity the diffusive contribution of the wire and thus treating it as a disordered low-resistance node connecting four terminals. The approximation in the second equality is made assuming the leads 1 and 2 to be much more resistive than the sinks: $G_1,G_2\ll G_A,G_B$. The coherent contribution (\ref{green}), on the other hand, is little affected by grounding the sinks since it is governed by the coherent electron propagation between the leads 1 and 2. Note that in the approximation of Eq.~(\ref{gsd}), $G_{12}$ is proportional to the product $G_1G_2$ which is of the same order in the tunneling amplitude as the coherent correction $\delta G_{12}$ given by Eq.~(\ref{green}).

In the rest of the paper, we will mainly focus on calculating the mesoscopic conductance (\ref{g}), which is sensitive to electron interference on traversing the distance $L$ between the leads. To that end, we will study the electronic structure in the long 1D wire as a function of parallel or perpendicular magnetic field for certain simple models.

In a clean 1D wire, $\mathcal{G}_i(L,E)\propto v_{gi}^{-1}$, the inverse group velocity, so that the conductance (\ref{g}) also appears sensitive to van Hove singularities. As we discuss in the following, however, disorder contributes to an exponential decay of the propagator $\mathcal{G}_i(L,E)$ near the band edge, which suppresses DOS singularities in the conductance (\ref{green}), making the interference effects more pronounced in the magnetoconductance traces.

It is interesting to compare Eq.~(\ref{green}) to the result obtained for a different physical situation by Fisher and Lee.\cite{fisherPRB81} These authors derive the Landauer-B\"{u}ttiker expression
\begin{equation}
G_w=\frac{e^2}{h}\sum_{ij}|t_{ij}|^2
\label{LB}
\end{equation}
for the conductance of a wire portion connecting two semi-infinite clean 1D reservoirs, using Kubo formula. Here $t_{ij}$ is the transmission-scattering amplitude between modes $i$ and $j$. The transmission coefficient through a clean region $t_i\propto v_{gi}\mathcal{G}_i$, for example, does not show any DOS singularities; in particular, $|t|^2=1$ independently of the density for propagating modes, giving the celebrated universal contribution to the conductance of $e^2/h$ per mode. The two-terminal conductance we are calculating cannot, in principle, exceed the conductance (\ref{LB}) of the wire itself. The van Hove singularities obtained in the tunneling approximation are, in fact, artifacts of the perturbative expansion in the tunneling amplitude, and must saturate at a value below the wire conductance (\ref{LB}). Note incidentally that in the diffuse limit $G_w$ corresponds to $G_L$ introduced in Eq.~(\ref{R12}).

\subsection{Cylindrical nanowires}

As a specific example, let us now consider a ballistic cylindrical wire with radius $r_0$. Because of the translational invariance and cylindrical symmetry of the wire, it is convenient to expand the one-electron wave function as
\begin{equation}
\hat{\varphi}(r,\theta,z)=e^{ikz}\sum_n\frac{e^{in\theta}}{\sqrt{r}}\hat{\varphi}_n(r)\,,
\label{phi}
\end{equation}
with a fixed momentum $k$ along the wire axis $z$. $\hat{\varphi}_n(r)$ is the radial-position-dependent spinor with a given orbital angular momentum $n$ along the wire. Assuming a uniform magnetic field, we can express the translationally invariant Hamiltonian $H$ according to
\begin{equation}
H\hat{\varphi}=e^{ikz}\sum_n\frac{e^{in\theta}}{\sqrt{r}}\sum_{n^\prime}\hat{H}_{n,n^\prime}\hat{\varphi}_{n^\prime}\,,
\end{equation}
in terms of the $2\times2$ matrices $\hat{H}_{n,n^\prime}$ in spin space mixing orbital angular momenta $n$ and $n^\prime$. For a quasi-1D wire with the electron radial-confinement potential $V(r)$, we obtain in a convenient gauge ($A_\theta=B_zr/2$ and $A_z=B_xy$)
\begin{align}
&2mr^2\hat{H}_{n,n^\prime}=[-r^2\partial_r^2+(kr)^2+(n+\phi_z)^2-1/4+U(r)\nonumber\\
&+g(m/m_e)\boldsymbol{\hat{\sigma}}\cdot\boldsymbol{\phi}]\delta_{n,n^\prime}+2ikr\phi_x(\delta_{n+1,n\prime}-\delta_{n-1,n^\prime})\nonumber\\
&-\phi_x^2(\delta_{n+2,n\prime}+\delta_{n-2,n^\prime}-2\delta_{nn^\prime})-mr\alpha_R\partial_rV(r)\nonumber\\
&\times\left[ikr\left(\hat{\sigma}_+\delta_{n+1,n^\prime}-\hat{\sigma}_-\delta_{n-1,n^\prime}\right)+2(n+\phi_z)\hat{\sigma}_z\delta_{n,n^\prime}\right.\nonumber\\
&\hspace{0.5cm}\left.+2\phi_x\hat{\sigma}_x\delta_{n,n^\prime}-\phi_x\left(\hat{\sigma}_+\delta_{n+2,n^\prime}+\hat{\sigma}_-\delta_{n-2,n^\prime}\right)\right]\,,
\label{H}
\end{align}
where $U(r)=2mr^2V(r)$ is a dimensionless confining potential, $\phi_{x,z}(r)=eB_{x,z}r^2/2$ is the respective magnetic-field flux (in  units of $h/e$) per area $\pi r^2$ [thus note the implied $r$ dependence of $\phi_{x,z}$ in Eq.~(\ref{H})], $-e$ is the electron charge, $m_e$ is the free-electron mass, $m$ is the effective electron mass, $g$ is the \textit{g}-factor, $\boldsymbol{\hat{\sigma}}=(\hat{\sigma}_x,\hat{\sigma}_y,\hat{\sigma}_z)$ is a vector of Pauli matrices, $\hat{\sigma}_{\pm}=\hat{\sigma}_x\pm i\hat{\sigma}_y$, and $B_{z(x)}$ is the magnetic-field component parallel (perpendicular) to the wire. Notice that the perpendicular field $B_x$ mixes angular-momentum states. The term proportional to $\alpha_R$ in Eq.~(\ref{H}) is due to the Rashba spin-orbit coupling in cylindrically-symmetric confinement $U(r)$ governed by the material-specific coefficient $\alpha_R$. Rashba interaction is believed to dominate spin-orbit coupling in InAs surface charge-accumulation layers due to a narrow gap and strong confinement fields.\cite{matsuyamaPRB00,schierholzPRB04,lamariPRB03,radantsevSEM03} The effective-mass approximation for the one-electron Hamiltonian (\ref{H}) breaks down near the wire surface, where the electron confinement $U(r)$ is sharp. Consequently, an appropriate boundary condition for the wave function has to be applied at the surface (here, we assume $\hat{\varphi}_n=0$ at the boundary), while $\partial_rU(r)$ defining the spin-orbit interaction is smooth within the electron-gas confinement.

\begin{figure}
\includegraphics[width=\linewidth,clip=]{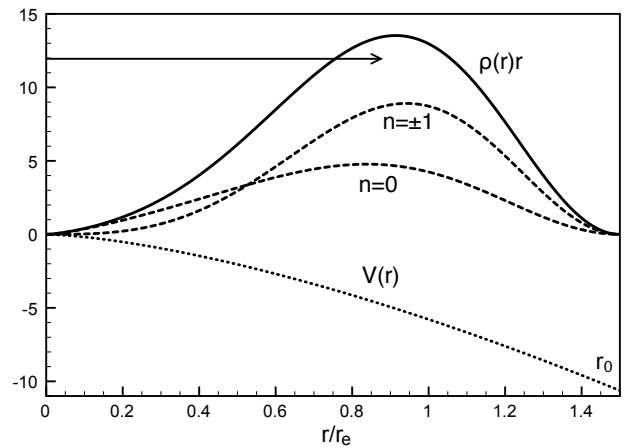}
\caption{\label{rho} Electron-density distribution in the radial direction at zero magnetic field and vanishing spin-orbit interaction. The solid line is the total density $\rho(r)$ multiplied by $r$ (in arbitrary units) and the dashed lines are respective contributions from the $n=0$ and $n=\pm1$ orbital angular-momentum states. The Fermi energy $E_F$ is chosen slightly below the $n=\pm2$ band edge. The confining potential $V(r)\propto -r^{3/2}$ is flat at the origin and has the slope $\beta$ near the hard-wall cutoff at $r=r_0$. The arrow shows the radius containing half of the electrons. The specific choice of the parameters $E_F$, $\beta$, and $r_0$ is discussed in the beginning of Sec.~\ref{results}. $r_e$ is a somewhat arbitrarily chosen ``effective radius" of the electron distribution (see Sec.~\ref{results} for more details).}
\end{figure}

Depending on the confining potential $V(r)$ and the wire radius, the radial electron-density distribution in the wire can be either peaked near the boundary or spread over the entire wire cross section. Figure~\ref{rho} shows the zero-field electron-density profile within the wire, calculated using parameters relevant to InAs nanowires (but for the purpose of demonstration setting $\alpha_R=0$). The $n=0$ mode has some finite density $\rho$ at the center of the wire [note that we are plotting $\rho(r)r$ in the figure], while the finite-$n$ modes do not contribute at the center and are pushed towards the boundary by the $n^2/2mr^2$ potential in the cylindrical coordinates [see Eq.~(\ref{H}].

The disorder-averaged Green's function for the $i$th mode is readily obtained after we calculate its dispersion $\epsilon_i(k)$ from Eq.~(\ref{H}) (assuming weak disorder),
\begin{align}
\mathcal{G}_i(L,E_F)&=\int_{-\infty}^{\infty}\frac{dke^{ikL}}{E_F-\epsilon_i(k)+i\Gamma}\nonumber\\
&=\sum_{{\rm Im}(k_i)>0}\frac{1}{v_{gi}(k_i)}e^{ik_iL}\,,
\label{G}
\end{align}
where the sum is over simple poles of $[E_F-\epsilon_i(k)+i\Gamma]^{-1}$ in the upper imaginary half plane [assuming no other singularities under the analytic continuation of $\epsilon_i(k)$ in the imaginary direction] and $v_{gi}$ is the corresponding group velocity [see Eq.~(\ref{vk})]. The disorder is accounted for in Eq.~(\ref{G}) by an imaginary self-energy $\Gamma=1/2\tau$ characterized by the scattering rate $1/\tau$. For weak disorder, we only take into account poles along the real axis with group velocity $v_{gi}$ at $\epsilon_i(k)=E_F$, the Fermi energy, and a corresponding imaginary contribution to the wave vector of
\begin{equation}
\delta k_i\approx i\Gamma/v_{gi}\,.
\label{dk}
\end{equation}
In particular, we see that close to van Hove singularities where $v_{gi}\to0$, the propagator (\ref{G}) is strongly damped due to disorder scattering. We will characterize the scattering strength by the mean free path $\lambda=v_F\tau$ using Fermi velocity $v_F$ of the two-dimensional (2D) electron gas corresponding to the surface charge layer.

Since we base the rest of our discussion on the form (\ref{G}) of the disorder-averaged retarded Green's function, which determines the conductance $\delta G_{12}$ through Eq.~(\ref{g}), it is important to summarize our physical assumptions underlying these approximations. In Eq.~(\ref{g}), we only keep a coherent ballistic contribution that decays exponentially with intra- and interband scattering (thus neglecting vertex corrections). In particular, we disregard terms that are affected by multiple scattering between disordered regions. This is a major limitation if, e.g., there are strong-scattering regions at the contacts induced by the lead-deposition procedure. We have assumed that injected electrons that move to the left and those electrons that have passed under the right lead do not contribute to $\delta G_{12}$. Finally, cross-sectional shape and size variation of the wire could affect the interference pattern due to the magnetic field. For example, a strong radial variation can deteriorate the periodicity and strength of the magneto-oscillations in the parallel field if the electrons are confined at the surface.

In summary, we study the conductance of translationally invariant nanowires, focusing on MC oscillations due to the ballistic paths between the leads. It is assumed that the paths that involve scattering from disorder result in rapidly oscillating contributions, which average out giving a nearly structureless background for the MC oscillations related to the ballistic interference (apart from the density-of-states singularities of the incoherent contribution $G_{12}$ to the conductance discussed in the previous subsection). If this is not the case, then our calculations would apply to the mean conductance for a set of nominally identical samples. Electron-electron interactions are not included, but are briefly commented on in Sec.~\ref{ee}.

\subsection{Cylindrical-shell model}
\label{csm}

As a further simplification, let us for a moment consider the shell model where electrons are confined in the radial direction to a narrow cylindrical well at radius $r_e$. Summing over the leads' degrees of freedom $\mu$, $\nu$ in Eq.~(\ref{g}) and assuming the leads are metallic with high electron density, we arrive at the following expression for the coherent conductance:
\begin{eqnarray}
\delta G_{12}&\propto&\sum_{i,j}\mathcal{G}_i\mathcal{G}_j^\ast(L,E_F)\int_{-\pi}^{\pi}d\theta|\lambda_s(\theta)|^2\hat{\varphi}^\dagger_{i}(\theta)\hat{\varphi}_{j}(\theta)\nonumber\\
&&\times\int_{-\pi}^{\pi}d\theta^\prime|\lambda_d(\theta^\prime)|^2\hat{\varphi}^\dagger_{j}(\theta^\prime)\hat{\varphi}_i(\theta^\prime)\,.
\end{eqnarray}
Here, $\hat{\varphi}_i(\theta)$ is the angular component of the electron spinor wave function at the Fermi energy in the $i$th mode of the wire and $\lambda_{s(d)}(\theta)$ is the tunneling amplitude at the source (drain) lead, which is assumed to be independent of the electronic state in the leads but angle-dependent around the wire circumference. In the following, we set the tunneling amplitude to be $\lambda_{s,d}(\theta)=1$ for $|\theta|<\theta_0$ and zero otherwise, for both contacts. Finally, expressing the wave functions in terms of the angular-momentum eigenstates, $\hat{\varphi}_i(\theta)=\sum_{n}\hat{c}_{in}e^{in\theta}$, we obtain
\begin{equation}
\delta G_{12}\propto\sum_{i,j}\mathcal{G}_i\mathcal{G}_j^\ast(L,E_F)\left|\sum_{nn^\prime}\hat{c}^\dagger_{in}\hat{c}_{jn^\prime}\frac{\sin[(n-n^\prime)\theta_0]}{(n-n^\prime)}\right|^2\,.
\label{sin}
\end{equation}

If $B_x=0$ and the spin-orbit coupling is neglected, $\alpha_R=0$, the $z$ component of the angular momentum commutes with the Hamiltonian (\ref{H}), and therefore $\hat{c}_{in}=|s\rangle\delta_{in}$, where $|s\rangle$ is the spin-$s$ state along the $z$ axis and $\delta_{in}$ is the Kronecker delta. If we also neglect the Zeeman coupling, then by gauge invariance, the conductance of our shell-model wire must be a periodic function of the flux $\phi_z=eB_zr_e^2/2$, so we may write $\delta G_{12}(\phi_z)=\sum_{m=-\infty}^\infty g_m\exp(i2\pi m\phi_z)$. However, there may be considerable weight in the higher harmonics of the fundamental frequency, so that $|g_m|$ may not be largest for small values of $m$. Also, the magnitude of $\delta G_{12}$ may oscillate multiple times within the period $-1/2<\phi_z<1/2$, and it is natural to ask whether there may be a characteristic frequency of such oscillations. Although there is no simple answer to this question, it is helpful to look at the physical content of the interference terms
\begin{equation}
G^{(nn^\prime)}=\mbox{Re}\mathcal{G}_n\mathcal{G}_{n^\prime}^\ast\left|\frac{\sin[(n-n^\prime)\theta_0]}{n-n^\prime}\right|^2
\label{gnn}
\end{equation}
between modes $n$ and $n^\prime$ arising in this idealized situation (contributing to the conductance $\delta G_{12}\propto\sum_{nn^\prime}G^{(nn^\prime)}$). The last factor in Eq.~(\ref{gnn}) describes the suppression of the interference between modes with different orbital angular momenta due to the finite injection and detection half-angle $\theta_0$. In particular, if $\theta_0=\pi$, all the cross terms vanish and $G^{(nn^\prime)}=|\mathcal{G}_n|^2\delta_{nn^\prime}$, reminiscent of the result\cite{fisherPRB81} for the Kubo conductance of a finite-length wire (\ref{LB}). Otherwise, we have to consider the oscillatory part of the cross terms proportional to $\mbox{Re}\mathcal{G}_n\mathcal{G}_{n^\prime}$: $G^{(nn^\prime)}\propto\cos[(k_n-k_{n^\prime})L]\exp[-\Gamma L(1/v_n+1/v_{n^\prime})]/v_nv_{n^\prime}$, where $k_n$ is the Fermi wave vector and $v_n$ group velocity for the $n$th mode along the wire. The corresponding rate of $G^{(nn^\prime)}$ phase accumulation in magnetic flux $\phi_z$ is given by
\begin{equation}
\omega_{nn^\prime}=\frac{L}{2\pi}\frac{\partial(k_n-k_{n^\prime})}{\partial\phi_z}\,.
\label{wnn}
\end{equation}
For the shell model with weak scattering, $k_n^2=2mE_F-(n+\phi_z)^2/r_e^2$ and, fixing $E_F$, $(L/2\pi)\partial k_n/\partial\phi_z=-(v_\theta/v_n)(L/2\pi r_e)$, where $mv_\theta=(n+\phi_z)/r_e$ and $mv_n=k_n$. $\omega_{nn^\prime}$ is thus equal to the relative winding number of the classical trajectories corresponding to modes $n$ and $n^\prime$, on traversing the wire of length $L$ around the cylinder of radius $r_e$. Since $\omega_{nn^\prime}$ is itself field dependent, it does not correspond to Fourier components of $G^{(nn^\prime)}(\phi_z)$, however. In addition, the latter is modulated by the magnetic field via the prefactor $\exp[-\Gamma L(1/v_n+1/v_{n^\prime})]/v_nv_{n^\prime}$ which has sharp features near the van Hove singularities in clean systems. Note that the fastest phase accumulation rates $\omega_{nn^\prime}$ correspond to small velocities $v_n$ along the wire and are thus more sensitive to disorder scattering.

We now consider the effect of a perpendicular field $B_x$ on the energy spectrum and tunneling DOS. We shall relax our assumption of a zero-thickness shell model, but we continue to assume that the confinement is sufficiently strong that at most one radial state is occupied for any given angular momentum $n$. Turning on the perpendicular field $B_x$ perturbatively gives the following correction to the energy eigenstate with angular momentum $n$ (disregarding spin-orbit interaction):
\begin{equation}
\Delta\epsilon_n(k)\approx\frac{(eB_x)^2}{4m}\left(\langle r^2\rangle_{nn}+\frac{k^2}{m}\sum_{n^\prime=n\pm1}\frac{\left|\langle r\rangle_{nn^\prime}\right|^2}{\epsilon_n-\epsilon_{n^\prime}}\right)\,,
\label{de}
\end{equation}
which is readily obtained by second-order perturbation theory using Eq.~(\ref{H}) [note that the lowest-order energy correction is quadratic in $B_x$ since the term linear in $B_x$ in the Hamiltonian (\ref{H}) mixes angular-momentum states that have different energies]. Here, $\langle r^2\rangle_{nn}$ is the expectation value of the radial position squared in the $n$th state and $\langle r\rangle_{nn^\prime}$ is the radial-position matrix element between \textit{radial} wave functions of states $n$ and $n^\prime$. For example, applying this result to the shell model ($\langle r^2\rangle_{nn}=|\langle r\rangle_{nn^\prime}|^2=r_e^2$) gives (omitting Zeeman energy)
\begin{equation}
\epsilon_n(k)\approx\frac{(n+\phi_z)^2+2\phi_x^2}{2mr_e^2}+\frac{k^2}{2m}\left[1+\frac{2\phi_x^2}{(n+\phi_z)^2-1/4}\right]\,,
\label{en}
\end{equation}
where $\phi_x=eB_xr_e^2/2$ is the flux that the perpendicular field $B_x$ would produce in an area $\pi r_e^2$, in units of the flux quantum. We see that for a wire of circular cross section, a weak magnetic field in the perpendicular direction has less effect on the energies than the same field along the wire, since the former produces energy shifts proportional to $\phi_x^2$ while the latter produces shifts linear in $\phi_z$, for $n\neq0$. The linear dependence on $\phi_z$ is a consequence of degeneracy between states of $n$ and $-n$, which will in general be lifted by any deviations from circular symmetry. The energy dependence on $\phi_z$ is then quadratic for $\phi_z\to0$, but the curvature will be large if the zero-field splitting between $n$ and $-n$ is small, and the dependence on $|\phi_z|$ becomes linear again when the shift due to $\phi_z$ is larger than the zero-field splitting.

\subsection{Electron-electron interactions}
\label{ee}

There can be several effects due to the electron-electron interactions affecting magnetoconductance oscillations in quasi-1D wires. In both disordered and clean wires, interactions lead to the inelastic-scattering rate growing with temperature. In the most naive form, this may be accounted for by adding a contribution to the scattering rate $\Gamma$ characterizing the single-electron propagator (\ref{G}). A finite temperature can also lead to a qualitatively different effect, even in the absence of inelastic scattering: The MC oscillation amplitude is suppressed as $T^{-1/2}$ when $k_BT>E_{\rm Th}$, the Thouless energy, i.e., the inverse time for electron transfer between the contacts.\cite{beenakkerRMP97}

For a wire that is connected to the outside world only through weak tunnel junctions, if the wire is short so that its Coulomb charging energy $E_C=e^2/2C$ is larger than the temperature, the usual Coulomb-blockade resonant structure for the incoherent (sequential-tunneling) conductance $G_{12}$ will set in as a function of source-drain and gate voltages. Away from the resonances, and at sufficiently low temperatures, higher-order quantum tunneling of the electric charge (i.e., the cotunneling) will dominate the current, see, e.g., Ref.~\onlinecite{averinPRL90}. For free electrons, the elastic component of the cotunneling signal reduces to our coherent conductance $\delta G_{12}$, Eq.~(\ref{green}). The Coulomb-blockaded regime in short wires thus opens an interesting venue for studying quantum-interference effects, which are higher-order in tunneling.

At low temperatures and vanishing disorder, another interesting regime may be achieved characteristic of the nanowire geometry: that of the Luttinger-liquid state of the 1D modes.\cite{voitRPP94} In that case, the tunneling amplitude acquires a power-law suppression at each contact with lowering temperature and bias,\cite{bockrathNAT99} determined by the anomalous compressibility of interacting electrons in 1D. Some qualitative features of the ballistic magneto-oscillation patterns, however, may remain similar to the noninteracting electrons. For example, the gauge invariance still dictates the $h/e$ periodicity in the parallel magnetic field for the narrow-confinement model, if we neglect the Zeeman energy. The spectral composition of the oscillations is, however, in general richer than discussed for noninteracting electrons in Sec.~\ref{csm}: In addition to the interference between one-electron trajectories between the contacts, there will also be contributions involving other electrons in the wire excited by the tunneling, which can similarly collect phase due to the applied magnetic field. Furthermore, the spin-charge separation characteristic of 1D electron liquids may profoundly effect the interference at a finite voltage bias.\cite{tserkovPRL02hw} These topics are, however, beyond the scope of the present paper.

\section{Numerical results}
\label{results}

As an application of the above formalism, we will here calculate the magnetoconductance in a model approximating InAs nanowires. InAs surface (or interfaces with other materials) exhibits unusual properties, which are not manifested by most other semiconductors. InAs interfaces with metals typically do not form Schottky barriers, making it possible to achieve {\O}hmic contacts (which was utilized, e.g., in Ref.~\onlinecite{dohSCI05} for using InAs nanowires as tunable supercurrent links between superconducting leads). Related to this is the property of InAs surfaces to support two-dimensional quantum-well states whose properties can be tuned by the bulk doping level.\cite{lamariPRB03} Owing to this surface-charge layer, 1D nanowires fabricated out of InAs can be expected to carry a finite number of 1D quantum channels even when the bulk is insulating.\cite{hansenPRB05,lieberPRIV02} For example, taking the 2D electron density to be $n_{2D}=10^{12}$~cm$^{-2}$,\cite{olssonPRL96,matsuyamaPRB00} we find that a 2D electron gas with radius $r_e\sim10$~nm can support as many as 10 quantum channels.

In specific calculations, we will take the parameters\cite{matsuyamaPRB00} measured for a 2D inversion layer at the surface of a \textit{p}-doped InAs bulk: Effective mass $m=0.027m_e$, \textit{g}-factor $g=-15$, Rashba coefficient $\alpha_R=1$~nm$^2$, disorder-scattering mean-free path $\lambda=100$~nm, and triangular-well confinement with the slope $\partial_rV=15$~meV/nm, which corresponds to the classical turning point of the ground state at the depth of about 10~nm. We will take the injection/detection half angle to be $\theta_0=\pi/5$, lead separation $L=300$~nm, and radius $r_0=15$~nm which, as explained below, corresponds to an effective radius of $r_e\sim10$~nm for electron distribution if using a quasi-2D shell model discussed in Sec.~\ref{csm} (assuming that only the lowest quantum-well state is occupied). The characteristic spin-orbit splitting $2\alpha_R\partial_rVk_F$ is of the order of $10$~meV. For simplicity, we will disregard the excited well states (which is reasonable for the chosen parameters), keeping in mind that the profile of the charge distribution can still be nontrivial on the scale of the wire radius. We fix the 1D electron density, regardless of magnetic field, to approximately correspond to a 2D electron gas with density $n_{2D}=10^{12}$~cm$^{-2}$ residing at radius $r_e$ (more specifically, the density is chosen so that the zero-field lowest-band Fermi wave vector along the wire is the same as that of the 2D electron gas with density $n_{2D}$).

\subsection{Narrow-well confinement (``shell model")}

Let us first consider an idealized situation of a narrow-well confinement of the electron distribution at radius $r_e$, which could be relevant for wider nanowires. As before, we define the (dimensionless) magnetic flux to be $\phi_{x(z)}=eB_{x(z)}r_e^2/2$. In the absence of the transverse field, $\phi_x=0$, and neglecting the Zeeman splitting, gauge invariance dictates that the conductance is periodic in $\phi_z$ with the period of unity. The periodicity is, however, broken by a finite \textit{g}-factor and by an extended confinement in the radial direction. According to the time-reversal symmetry, the linear conductance is symmetric under magnetic-field inversion, $G(-B)=G(B)$, so that we need to consider positive fields only.

\begin{figure}
\includegraphics[width=\linewidth,clip=]{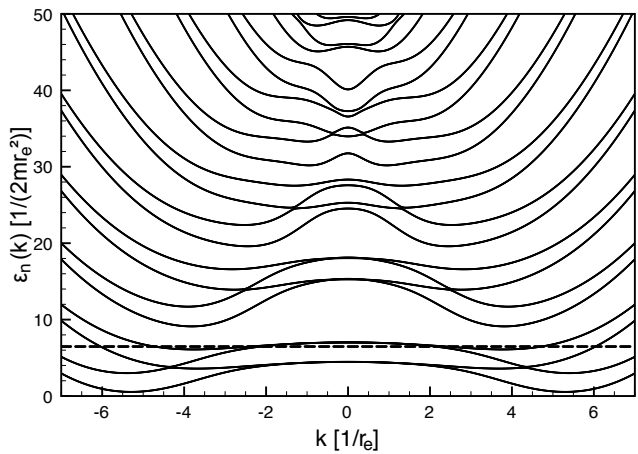}
\caption{\label{Ekx} Band structure of the 1D wire in a large perpendicular magnetic field, $\phi_x=3$ ($\phi_z=0$), for the shell model with $r_e=10$~nm. Following a complicated nonmonotonic behavior of $\epsilon_n(k)$ at small $k$ and $n$, the energy grows nearly linearly with $k$ at larger $k$, and eventually with the usual $\propto k^2$ dispersion at $k\rightarrow\infty$. These different regimes correspond to the dominance of the terms on the right-hand side of Eq.~(\ref{H}) having the respective scaling with $k$. The dashed line shows the Fermi-level position.}
\end{figure}

\begin{figure}
\includegraphics[width=\linewidth,clip=]{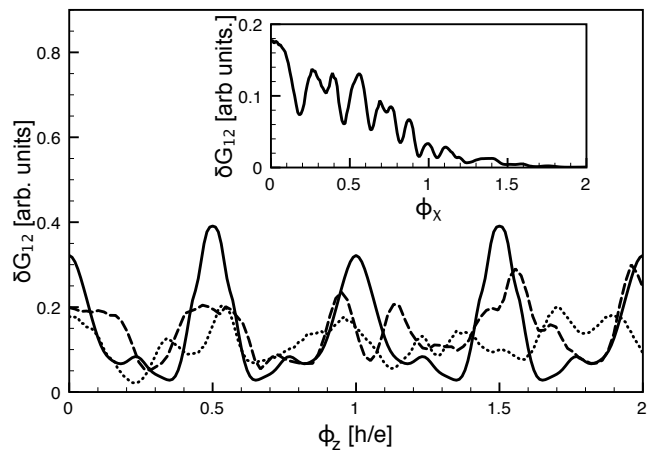}
\caption{\label{fft}The main panel shows the interference MC traces in the parallel magnetic field for an idealized shell model with $r_e=10$~nm, $g=0$, and $\alpha_R=0$ (solid line), a shell model including Zeeman splitting and spin-orbit coupling (dashed line), and a triangular-well model with outer wire radius $r_0=15$~nm (dotted line). All other parameters are listed in the text (in particular, $L=300$~nm). The inset shows a calculation similar to the dotted line in the main panel but for the perpendicular magnetic field. The quantum-well states are treated in the WKB approximation (see Sec.~\ref{twc} for details).}
\end{figure}

The 1D dispersion is given by Eq.~(\ref{en}) at low perpendicular fields, $\phi_x\ll1$, and vanishing spin-orbit coupling, $\alpha_R=0$. In general, we can calculate the dispersion numerically by diagonalizing Hamiltonian (\ref{H}) (see Fig.~\ref{Ekx} for $\phi_x=3$). In order to find the conductance at a finite magnetic field, we first calculate the Fermi level  keeping electron density fixed and then find the crossing points of $\epsilon_n(k)=E_F$ with positive slopes. The Green's functions for corresponding modes are then evaluated using Eq.~(\ref{G}) and the conductance is finally obtained according to Eq.~(\ref{sin}), which requires knowing wave functions as well as dispersions [both given by the diagonalization of Eq.~(\ref{H})]. Examples of the magnetoconductance traces are shown in Fig.~\ref{fft} for parallel and perpendicular field orientations, taking into account both Zeeman splitting and spin-orbit coupling. In the case of the parallel field, the conductance is in general aperiodic in $\phi_z$, although some reminiscence of the $h/e$ periodicity is still present.

\begin{figure}
\includegraphics[width=\linewidth,clip=]{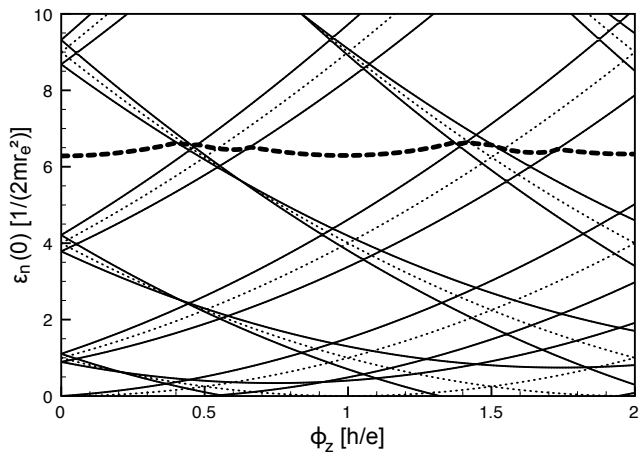}
\caption{\label{Ez} $k=0$ energies for quasi-1D bands calculated by diagonalizing Hamiltonian~(\ref{H}) with the magnetic field along the wire axis for a shell model with $r_e=10$~nm. Dotted lines show spin-degenerate energies after setting $g=0$ and $\alpha_R=0$. Solid lines are calculated taking into account the Zeeman and spin-orbit energies. Zero-field spin degeneracy is lifted by the latter while Zeeman splitting dominates at $\phi_z\gg1$. The dashed line is the Fermi energy corresponding to the solid-line energies, keeping the total 1D electron density fixed.}
\end{figure}

In the Fig.~\ref{fft} inset, there are considerable MC oscillations at low perpendicular fields, before the conductance is suppressed at larger fields. The latter can be qualitatively understood to be due to the onset of the Landau-level formation in the perpendicular magnetic field, which flattens the dispersions (see Fig.~\ref{Ekx}), lowers group velocity along the wire, and therefore, according to Eq.~(\ref{dk}), enhances disorder scattering on the propagation between the contacts. It is worthwhile noting in this regard one general feature near the van Hove singularities: Fixing the 1D electron density, the Fermi level tends to be somewhat attracted to the van Hove singularities before 1D modes disappear or after they start getting populated, due to the diverging compressibility. The corresponding features can be seen in Fig.~\ref{Ez} in the Fermi-level position close to where it crosses transverse energy levels in the parallel magnetic field.

\begin{figure}
\includegraphics[width=\linewidth,clip=]{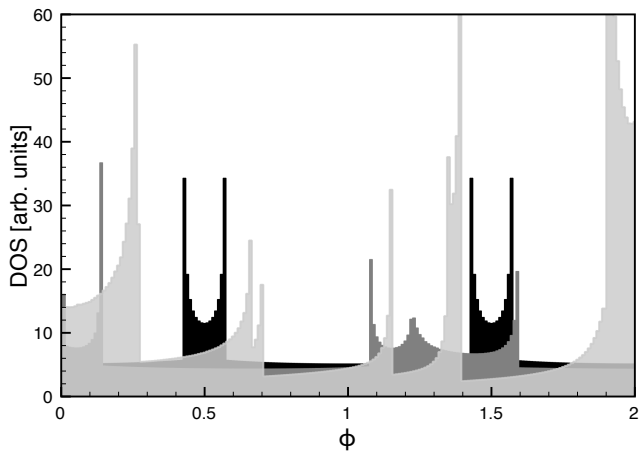}
\caption{\label{DOS} The total density of states as a function of the parallel (black and dark-gray histograms) or perpendicular magnetic field (light-gray histogram), which governs the incoherent two-terminal double-tunnel-barrier conductance (\ref{R12}) according to Eq.~(\ref{gk}). The black histogram was calculated using the same parameters as the solid line in the main panel of Fig.~\ref{fft}, the dark-gray histogram|the dashed line, and the light-gray histogram|the dotted line.}
\end{figure}

As discussed in Sec.~\ref{gc}, see Eqs.~(\ref{R12}) and (\ref{gk}), the two-terminal source-drain magnetoconductance is expected to be dominated by the density-of-states singularities. It is plotted in Fig.~\ref{DOS}. By comparing to Fig.~\ref{fft}, we see that the sharp DOS features are fully suppressed in the coherent-tunneling contribution to the conductance due to the significant disorder scattering. We note that if one was able to identify the positions of these van Hove singularities in the two-terminal measurements or by other means, as a function of magnetic fields $B_z$ and $B_x$, this would, in principle, allow us to obtain a wealth of information about the band structure and the electron distribution in the radial direction [by using relations like Eq.~(\ref{de}) obtained for low perpendicular fields $B_x$]. A detailed procedure would require a further theoretical analysis.

\begin{figure}
\includegraphics[width=\linewidth,clip=]{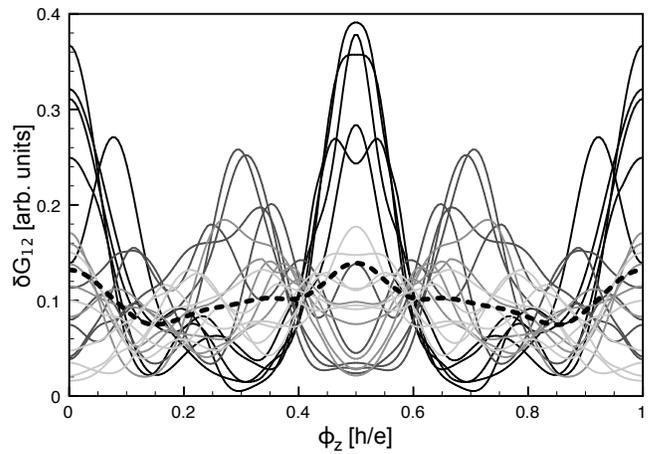}
\caption{\label{int}Shell-model calculation of the interference magnetoconductance for 2D electron densities between $1$ and $3$ in $0.1$ increments (with lighter line shades corresponding to higher densities), in units of $10^{12}~{\rm cm}^{-2}$. The dashed line is the average. Here, $g$ and $\alpha_R$ are set to zero, so that the $1\times10^{12}$~cm$^{-2}$ trace is the same as the solid line in Fig.~\ref{fft}.}
\end{figure}

Finally, before closing this subsection, we would like to note an interesting peculiarity of the magnetoconductance at low parallel magnetic fields. Varying the electron density, we find that there is more often a maximum of $\delta G_{12}(\phi_z)$ at $\phi_z=0$, rather than a minimum (there must be one of the two due to the time-reversal symmetry), over a wide range of parameters. This ballistic-interference maximum at low fields persists over density averaging, but it is different from the weak-antilocalization peak in the presence of a strong spin-orbit interaction in diffuse wires. See Fig.~\ref{int} for the shell-model calculation with zero spin-orbit coupling and Zeeman splitting, where the 2D electron density is varied between $1$ and $3$ in $0.1$ increments, in units of $10^{12}~{\rm cm}^{-2}$. We, however, expect the opposite statistical property, i.e., a local minimum at zero field for the incoherent contribution to the conductance $G_{12}$.

\subsection{Triangular-well confinement}
\label{twc}

\begin{figure}
\includegraphics[width=\linewidth,clip=]{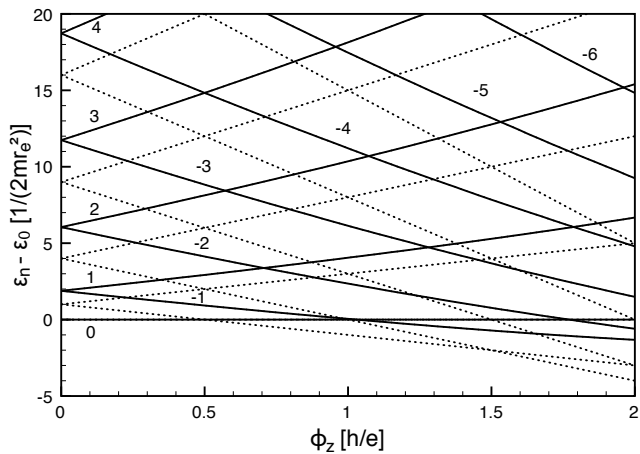}
\caption{\label{EWKB} Solid lines are the ground-state energies (relative to the $n=0$ energy) for transverse triangular-well states with fixed angular momenta $n$ (labeled in the plot) as a function of the parallel magnetic field, setting $g=0$ and $\alpha_R=0$. The wire radius $r_0=15$~nm and other parameters are listed in the beginning of Sec.~\ref{results}. Dotted lines are calculated for the shell model with $r_e=10$~nm, which is a characteristic radius for the radial-distribution center of mass at small angular momenta $n$. See Fig.~\ref{rho} for a similar confining potential $V(r)$ (although not strictly triangular as here). $\phi_{z}=eB_{z}r_e^2/2$.}
\end{figure}

Consider now a triangular confinement potential
\begin{equation}
V(r)=\beta(r_0-r)+\infty\Theta(r-r_0)\,,
\end{equation}
where $\beta$ is the confinement steepness and $\Theta$ is the Heaviside step function defining the sharp cutoff at the outer radius $r_0$. For the magnetic field along the wire axis, the total effective potential in the radial direction depends on the angular-momentum state, the higher momenta (with respect to the flux $-\phi_z$) being pushed more outward, as can be seen from Eq.~(\ref{H}). We calculate the corresponding transverse ground-state energies for different angular momenta using the WKB approximation, in the absence of the spin-orbit interaction and perpendicular magnetic field. For simplicity, we use Eq.~(\ref{gnn}) for suppression of the cross terms describing the interference between different angular-momentum states, which was derived for the shell model. The spin-orbit coupling and, if present, the perpendicular magnetic field are also included similarly to the shell model, after we choose an appropriate effective radius $r_e$ [by comparing triangular-well and shell-model energies (see Fig.~\ref{EWKB})]. The calculated magnetoconductance for the parallel magnetic field is plotted in Fig.~\ref{fft} by the dotted line.

In Fig.~\ref{EWKB}, we plot the parallel-field dependence of the triangular-well ground-state energies at fixed angular momenta $n$, which were used in calculating the magnetoconductance. In order to compare WKB and shell-model calculations, the lowest energy is set to zero. For larger $n$ (with respect to $-\phi_z$), the radial potential in Eq.~(\ref{H}) pushes electrons outwards so that the electrons become confined to the narrow well near the surface at $r=r_0$, as $n\rightarrow\infty$, decreasing the period of magneto-oscillations. This trend can be seen in Fig.~\ref{EWKB}. In particular, this means that it is rather crude to characterize different angular-momentum states by a single effective radius $r_e$ and one would have to solve Eq.~(\ref{H}) for a fully three-dimensional electron distribution (and beyond the WKB approximation, which is not expected to be very accurate for quantum-well ground-state energies in any case) if precise energy levels are desired. The zero-field radial electron distribution is shown in Fig.~\ref{rho} for a somewhat more realistic confinement potential, which is flat at the origin and normalized to have slope $\beta$ at the boundary. The electron distribution is indeed rather broad within the wire. For a quantitative analysis, it would be necessary to determine the confining potential $V(r)$ self-consistently with the electron distribution, as a function of the magnetic field.

\section{Discussion}
\label{summary}

We theoretically studied the magnetoconductance (MC) of cylindrical quasi-one-dimensional nanowires carrying several quantum channels confined near the surface. We focused on the MC oscillations arising due to the ballistic interference of electrons propagating via several 1D modes along a finite-length wire section between a metallic source and drain leads. Van Hove singularities in the 1D density of states are expected to dominate most prominent MC features in a two-terminal configuration, and we have therefore suggested to use additional reservoirs at the ends of the wire to absorb injected electrons, which diffused away from the leads, in order to enhance the more interesting ballistic-interference features.

In order to calculate mesoscopic MC interference patterns, we have studied the electronic structure of 1D modes in the long wires for an arbitrary uniform magnetic field, taking into account Rashba spin-orbit coupling due to the surface-confinement potential. In the case of the parallel magnetic field along the wire axis $z$, the MC is periodic in the magnetic flux $\phi_z$ with a period of $h/e$ only in the case of a narrow surface confinement and vanishing Zeeman splitting. Relaxing these assumptions, the conductance oscillations become aperiodic but some remnants of periodicity are still visible for realistic parameters, as seen in Fig.~\ref{fft}. A large perpendicular magnetic field leads to the Landau-level formation and suppression of the discussed mesoscopic magnetoconductance. The amplitude of the conductance fluctuations is reduced at a finite voltage bias and temperature or upon gate-voltage averaging modulating electron density. Certain interference features, such as the local maximum in $\delta G_{12}$ at zero parallel magnetic field, however, can survive energy averaging (see Fig.~\ref{int}).

\begin{acknowledgments}
The discussion presented here was motivated by the experiments on the magnetoconductance of InAs nanowires performed by A. Greytak, K. Kim, and C. M. Lieber at Harvard University. This work was supported in part by NSF Grants No. PHY 01-17795 and No. DMR 05-41988 and by the Harvard Society of Fellows.
\end{acknowledgments}

\end{document}